# Varying physical constants and the lithium problem

Rajendra P. Gupta*

*Department of Physics, University of Ottawa, Ottawa, Canada K1N 6N5*

We have used the recently published varying physical constants (VPC) approach in an attempt to resolve the primordial lithium abundance problem. The value of the ratio of $^7$Li to hydrogen $^7$Li/H = $1.400\ (\pm 0.023) \times 10^{-10}$ we have calculated using this approach is about four times lower than that estimated using the standard lambda cold dark matter (ΛCDM) cosmological model, and is consistent with the most agreed observational value of $1.6\ (\pm 0.3) \times 10^{-10}$. In the VPC approach Einstein equations are modified to include the variation of the speed of light $c$, gravitational constant $G$ and cosmological constant Λ using the Einstein-Hilbert action. Application of this approach to cosmology naturally leads to the variation of the reduced Plank constant $\hbar$ and the Boltzmann constant $k_B$ as well. They approach fixed values at the scale factor $a \ll 1$: $c = c_0/e$, $G = G_0/e^3$, $\hbar = \hbar_0/e$ and $k_B = k_{B0}/e^{5/4}$, where $e$ is the Euler's number ($= 2.7183$). Since the VPC cosmology reduces to the same *form* as the ΛCDM cosmology at very small scale factors, we could use an existing Big-Bang nucleosynthesis (BBN) code AlterBBN with the above changes to calculate the light element abundances under the VPC cosmology. Among other abundances we have calculated at baryon to photon ratio $\eta = 6.1 \times 10^{-10}$ are: $^4$He/H = $0.2478\ (\pm 0.0041)$, D/H = $2.453\ (\pm 0.041) \times 10^{-5}$, $^3$He/H = $2.940\ (\pm 0.049) \times 10^{-5}$.

## I. INTRODUCTION

The standard model has been immensely successful in explaining cosmological observations better than any of the alternatives offered from time to time. However, it is not able to resolve the so called 'lithium problem' [e.g. 1-4]: a gaping discrepancy exists between the observed primordial lithium abundance and that predicted by the standard model's Big-Bang nucleosynthesis (BBN). The subject has been extensively discussed in the literature. Here we cite the most recent reviews [5,6] which have exhaustive citations on BBN in general and the lithium problem in particular. Many astronomers consider the problem acute enough to suggest the existence of a new physics [e.g. 2,4,5,7,8].

Many alternative solutions for the lithium problem have been offered. Mathews et al. in 2019 [9] have succinctly reviewed them. The offered solutions include: nuclear and diffusion processes destroying lithium *after* BBN [10,11]; nuclear – reactions destroying lithium *during* BBN [12]; and cosmological – new physics beyond the standard BBN approach. Their focus was to overview a variety of possible cosmological solutions, and their shortcomings. The possibility of physical process that may modify the velocity distribution of particles from that given by Maxwell-Boltzmann statistics has also been considered [13,14]. The modification may be due to inhomogeneous spatial distribution of domains of primordial magnetic field strength [15]. Such modified velocity distribution could lead to reduced lithium abundance. The ambipolar diffusion of Li+ ions induced by magnetic field working after the BBN could also have contributed to the reduction in observed lithium abundance [74]. Another possibility is that scattering with the mildly relativistic electrons in the background plasma alters the baryon distribution to one resembling a Fermi-Dirac distribution [16,17], although on a closer scrutiny authors of reference [16] found that the distribution follows the Maxwellian distribution [75]. However, the models that are able to reduce the $^7$Li production invariably overproduce D. Hybrid models have been developed to resolve the resulting D/H problem as well as the $^7$Li/H. However, the hybrid models result in over producing $^6$Li; since the current upper limit on $^6$Li abundance is very loose it is difficult to exclude a model that predicts large $^6$Li/H. Mathews et al. [9], therefore, concluded that none of those alternatives are able to satisfactorily resolve the lithium problem.

In this paper our attempt is to resolve the lithium problem using the variable physical constant approach. The concept of varying physical constants has been in existence since 1883 as we know it [18-20]. It got traction when Dirac in 1937 [21,22] suggested the variation of the constant $G$ based on his large number hypothesis. Later, Brans and Dicke in 1961 [23] developed the $G$ variation theory in which constant $G$ was raised to the status of scalar field potential. Even Einstein, who developed his ground breaking theory of special relativity based on the constancy of the speed of light, considered its possible variation [24]. The varying speed of light theories were developed comprehensively by Dicke [25], Petit [26] and Moffatt [27,28]. Albrecht and Magueijo [29] and Barrow [30] developed such a theory in which Lorentz invariance is broken as there is a preferred frame in which scalar field is minimally coupled to gravity. Other theories include locally invariant theories [31,32] and vector field theories that cause spontaneous violation of Lorentz invariance [33].

A comprehensive review of the varying fundamental physical constants was undertaken by Uzan in 2003 [34] followed by his more recent review [35]. An update of the observational and experimental status of the constancy of physical constants was done by Chiba in 2011 [36].

Several attempts have been made to resolve the lithium problem by varying some constants. The variation of fine structure constant in BBN was considered by Ichikawa and Kawasaki [37], who also included the non-standard expansion rate of the Universe, and most recently by Clara and Martins [4] who determined that the fine structure constant was larger at the time of BBN than it is now. Landau et al. [38] attempted to resolve the problem by assuming that the gauge coupling constants were different during the Big-Bang era than they are now and thus affected the nuclear reaction rates. Dmitriev et al. [39] tried the variation in the deuterium binding energy. Coc et al. [40] considered the variation of Yukawa couplings and the fine structure constant as well as the variation of the deuterium binding energy. Berengut et al. [41] considered the effect of quark mass variation on big bang nucleosynthesis. Mori and Kusakabe [76] investigated the effects of a quark mass variation on the cross section of nuclear reactions and showed that the S-factor involved can sensitively change, especially for resonant or sub-threshold reactions.

In most of the proposed theories variable physical constants (VPC) are introduced at the cost of either not conserving energy-momentum or violating Bianchi identities. This leads to breaking the covariance of the theory. Such theories are considered inconsistent or *ad hoc* [42] or quasi-phenomenological [43]. One requires action principle to take into account the variation of the fundamental constants that are being considered for generalization of Einstein equations. This approach was attempted recently by Costa et al. [44,45] by considering the speed of light $c$, the gravitational constant $G$, and the cosmological constant $\Lambda$ as scalar fields. They introduced the Einstein-Hilbert action that is considered consistent with the Einstein equations, and the general constraint that is compliant with contracted Bianchi identities and standard local conservation laws. It preserves the invariance of the general relativity and thus is general covariant. The approach of Costa et al. has confirmed the findings of our quasi-phenomenological model [43].

*rgupta4@uottawa.ca

Based on the above, we recently were able to developed a VPC model and show [46] that the model: (a) fits the supernovae 1a observational data marginally better than the $\Lambda$CDM model; (b) determines the first peak in the power spectrum of the cosmic microwave background temperature anisotropies at multipole value of $l = 217.3$; (c) calculates the age of the Universe as 14.1 Gyr; and (d) finds the BAO acoustic scale to be 145.2 Mpc. These numbers are within a couple of percent of the accepted values. This success of the model encouraged us to consider applying the model to BBN as well.

We begin with establishing the theoretical background for our work in Section II. We will confine ourselves to consider only what needs to be modified in an existing, well established BBN code, developed for the standard $\Lambda$CDM model, in order to make it compliant with our VPC model.

Section III presents the results that we have obtained by modifying the well-known AlterBBN code [47] that we found relatively easy to modify as compared to some other codes we considered, for example PRIMAT [48]. In Section IV we discuss the findings of this paper and in Section V we present our conclusions.

## 2. THEORETICAL BACKGROUND

Following Costa et al. (2019) we may write the Einstein equations with varying physical constants with respect to time $t$ – the speed of light $c = c(t)$, the gravitational constant $G = G(t)$ and the cosmological constant $\Lambda = \Lambda(t)$ - applicable to the homogeneous and isotropic universe, as follows:

$$G_{\mu\nu} = \left(\frac{8\pi G(t)}{c(t)^4}\right) T_{\mu\nu} - \Lambda(t) g_{\mu\nu}. \quad (1)$$

Here $G_{\mu\nu} = R_{\mu\nu} - \frac{1}{2} g_{\mu\nu} R$ is the Einstein tensor with $R_{\mu\nu}$ the Ricci tensor and $R$ the Ricci scalar, and $T_{\mu\nu}$ is the stress energy tensor. Applying the contracted Bianchi identities, torsion free continuity and local conservation laws

$$\nabla^\mu G_{\mu\nu} = 0 \text{ and } \nabla^\mu T^{\mu\nu} = 0, \quad (2)$$

one gets a general constraint equation for the variation of the physical constants

$$\left[\frac{1}{G} \partial_\mu G - \frac{4}{c} \partial_\mu c\right] \left(\frac{8\pi G}{c^4}\right) T^{\mu\nu} - (\partial_\mu \Lambda) g^{\mu\nu} = 0. \quad (3)$$

Now the FLRW (Friedmann–Lemaître–Robertson–Walker) metric for the geometry of the universe is written as:

$$ds^2 = -c^2(t) dt^2 + a^2(t) \left[\frac{1}{1-kr^2} dr^2 + r^2(d\theta^2 + \sin^2\theta d\phi^2)\right], \quad (4)$$

with $k = -1, 0, +1$ depending on the spatial geometry of the universe: $-1$ for negatively curved universe, 0 for flat universe, and $+1$ for positively curved universe.

The stress-energy tensor, assuming that the universe contents can be treated as perfect fluid, is written as:

$$T^{\mu\nu} = \frac{1}{c^2(t)} (\varepsilon + p) U^\mu U^\nu + p g^{\mu\nu}. \quad (5)$$

Here $\varepsilon$ is the energy density, $p$ is the pressure, $U^\mu$ is the 4-velocity vector with the constraint $g_{\mu\nu} U^\mu U^\nu = -c^2(t)$. (Unless necessary to avoid confusion, we will drop showing $t$ variation, e.g. $c(t)$ is written as $c$.)

Solving the Einstein equation (such as by using Maple 2019) [49] then yields VPC compliant Friedmann equations:

$$H^2 \equiv \frac{\dot{a}^2}{a^2} = \frac{8\pi G \varepsilon}{3c^2} + \frac{\Lambda c^2}{3} - \frac{kc^2}{a^2}, \Rightarrow \dot{a}^2 = a^2 \left(\frac{8\pi G \varepsilon}{3c^2} + \frac{\Lambda c^2}{3} - \frac{kc^2}{a^2}\right), \quad (6)$$

$$\frac{\ddot{a}}{a} = -\frac{4\pi G}{3c^2}(\varepsilon + 3p) + \frac{\Lambda c^2}{3} + \frac{\dot{c}}{c} \frac{\dot{a}}{a} = -\frac{4\pi G}{3c^2}(\varepsilon + 3p) + \frac{\Lambda c^2}{3} + \frac{\dot{c}}{c} \frac{a}{\dot{a}} H^2. \quad (7)$$

Here a dot on top of a variable denotes the time derivative of that variable, e.g. $\dot{c} \equiv dc/dt$. Taking time derivative of Eq. (6), dividing by $2a\dot{a}$ and equating it with Eq. (7), yields the general continuity equation:

$$\dot{\varepsilon} + 3 \frac{\dot{a}}{a}(\varepsilon + p) = -\left[\left(\frac{\dot{G}}{G} - 4\frac{\dot{c}}{c}\right)\varepsilon + \frac{c^4}{8\pi G} \dot{\Lambda}\right] \quad (8)$$

Eq. (3) for the FLRW metric and perfect fluid stress-energy tensor reduces to:

$$\left[\left(\frac{\dot{G}}{G} - 4\frac{\dot{c}}{c}\right) \frac{8\pi G}{c^4} \varepsilon + \dot{\Lambda}\right] = 0, \quad (9)$$

therefore,

$$\dot{\varepsilon} + 3\frac{\dot{a}}{a}(\varepsilon + p) = 0. \quad (10)$$

Using the equation of state relation $p = w\varepsilon$ with $w = 0$ for matter and $w = 1/3$ for relativistic particles, the solution for this equation is $\varepsilon = \varepsilon_0 a^{-3-3w}$, where $\varepsilon_0$ is the current energy density of all the components of the universe when $a = a_0 = 1$.

Next we need to consider the continuity equation Eq. (9). When $\Lambda$ is constant, $\dot{G}/G = 4\dot{c}/c$. However, one could choose any relationship between $G$ and $c$, say $\dot{G}/G = \sigma \dot{c}/c$. Then from Eq. (9), by defining $\varepsilon_\Lambda = c^4 \Lambda/(8\pi G)$, we have

$$\frac{8\pi G}{c^4} \frac{\dot{c}}{c} (4 - \sigma) \varepsilon = \dot{\Lambda}, \Rightarrow \frac{\dot{c}}{c} = \frac{c^4 \Lambda}{8\pi G} \left(\frac{\dot{\Lambda}}{\Lambda}\right) \left(\frac{1}{(4-\sigma)\varepsilon}\right) \equiv \frac{\varepsilon_\Lambda}{(4-\sigma)\varepsilon} \frac{\dot{\Lambda}}{\Lambda},$$
$$\Rightarrow \varepsilon_\Lambda = \frac{\dot{c}}{c} \frac{\Lambda}{\dot{\Lambda}} (4 - \sigma) \varepsilon \quad (11)$$

The parameter $\sigma$ may be determined based on the physics or by fitting the observations. We have determined in the past [50] that $\sigma = 3$ analytically, i.e. $\dot{G}/G = 3\dot{c}/c$ and confirmed it by fitting the supernovae (SNe) 1a data [43,46]. Thus, we must have $\varepsilon_\Lambda = \dot{c} \Lambda \varepsilon/(c \dot{\Lambda})$.

The most common way of defining the variation of the constant is by using the scale factor powerlaw [51,52] such as $c = c_0 a^\alpha$ which results in $\dot{c}/c = \alpha \dot{a}/a = \alpha H$. The advantage is that it results in very simple Freedmann equations. However, as $a \to 0$ the variable constant tends to zero or infinity depending on the sign of $\alpha$. So, it yields reasonable results when $a = 1/(1 + z)$ corresponds to relatively small redshift $z$, but not for large $z$. We therefore have tried another relation that results in:

$$c = c_0 \exp[(a^\alpha - 1)]; \; G = G_0 \exp[3(a^\alpha - 1)]; \text{ and } \\ \Lambda = \Lambda_0 \exp[(a^\beta - 1)]. \tag{12}$$

Their limitation is that $c$ can decrease in the past at most by a factor of $e = 2.7183$ (the Euler number) and $G$ can decrease by a factor of $e^{-3}$ (for positive $\alpha$ within the region of their applicability). Using relations of Eq. (12), we can now write Eq. (11) for $\sigma = 3$,

$$\frac{c^4 \Lambda}{8\pi G} \equiv \varepsilon_\Lambda = \frac{\dot c}{c}\frac{\Lambda}{\dot\Lambda}\varepsilon = \frac{\alpha}{\beta}a^{\alpha-\beta}\varepsilon. \tag{13}$$

The first Friedmann equation, Eq. (6), becomes

$$H^2 = \frac{8\pi G}{3c^2}\left(\varepsilon + \frac{\Lambda c^4}{8\pi G}\right) - \frac{kc^2}{a^2} = \frac{8\pi G}{3c^2}\varepsilon\left(1 + \frac{\alpha}{\beta}a^{\alpha-\beta}\right) - \frac{kc^2}{a^2}. \tag{14}$$

Here energy density $\varepsilon = \varepsilon_m + \varepsilon_r = \varepsilon_{m,0}a^{-3} + \varepsilon_{r,0}a^{-4}$ with subscript $m$ for matter and $r$ for radiation (relativistic particles, e. g. photons and neutrinos). Dividing by $H_0^2$, we get

$$\frac{H^2}{H_0^2} = \frac{8\pi G}{3c^2 H_0^2}\left(\varepsilon_{m,0}a^{-3} + \varepsilon_{r,0}a^{-4}\right)\left(1 + \frac{\alpha}{\beta}a^{\alpha-\beta}\right) - \frac{kc^2}{H_0^2 a^2}. \tag{15}$$

At $t = t_0$ (current time), $a = 1$ and $H = H_0$. Therefore,

$$1 = \frac{8\pi G_0}{3c_0^2 H_0^2}(\varepsilon_{m,0} + \varepsilon_{r,0})\left(1 + \frac{\alpha}{\beta}\right) - \frac{kc_0^2}{H_0^2} \\ = (\Omega_{m,0} + \Omega_{r,0})\left(1 + \frac{\alpha}{\beta}\right) - \frac{kc_0^2}{H_0^2}. \tag{16}$$

Here we have defined the current critical density as $\varepsilon_{c,0} = 3c_0^2 H_0^2/8\pi G_0$, $\Omega_{m,0} = \varepsilon_{m,0}/\varepsilon_{c,0}$ and $\Omega_{r,0} = \varepsilon_{r,0}/\varepsilon_{c,0}$. Thus, by defining $\Omega_0 = (\Omega_{m,0} + \Omega_{r,0})(1 + \alpha/\beta)$, we may write Eq. (16)

$$\Omega_{k,0} \equiv -\frac{kc_0^2}{H_0^2} = 1 - \Omega_0, \text{ and} \tag{17}$$

$$\frac{H^2}{H_0^2} = \exp[(a^\alpha - 1)]\left(\Omega_{m,0}a^{-3} + \Omega_{r,0}a^{-4}\right)\left(1 + \frac{\alpha}{\beta}a^{\alpha-\beta}\right) + \\ \Omega_{k,0}\exp[2(a^\alpha - 1)]a^{-2}. \tag{18}$$

Now we have determined $\alpha = 1.8 = -\beta$ [46]. As $a \to 0$, applicable for the nucleosynthesis epoch, $\exp[(a^\alpha - 1)] \to 1/e$, and Eq. (18) approaches

$$\frac{H^2}{H_0^2} = \frac{1}{e}\left(\Omega_{m,0}a^{-3} + \Omega_{r,0}a^{-4}\right). \tag{19}$$

This equation is the same as for the $\Lambda$CDM model except for the factor $1/e$ that effectively alters the matter and radiation densities. But since both the densities are reduced by the same factor, the baryon to photon ratio $\eta$ remains the same for the VPC model and the $\Lambda$CDM model.

We can see that for $a \ll 1$, $c = c_0/e$ and $G = G_0/e^3$. We had also shown [46] that the reduced Planck constant $\hbar$ evolves as $\hbar_0 \exp[(a^\alpha - 1)]$ and the Boltzmann $k_B$ evolves as $k_{B,0}\exp[(5/4)(a^\alpha - 1)]$, which become $\hbar = \hbar_0/e$ and $k_B = k_{B,0}/e^{5/4}$ for $a \ll 1$. This leads to the conclusion that we can use any proven BBN code provided we meticulously redefine all the equations and parameters that may contain these constants. This is a nontrivial task as all codes use mixed units. Since we will be using the AlterBBN code [47,53], we refer reader to these references and citations therein for the theory used in the code. Nevertheless, we will consider specificities unique to VPC models.

Since predictions from any BBN model rely strongly on nuclear reaction rates and neutron lifetime measured *at the present time*, it is important to establish whether or not they have the same value back at the time when BBN took place.

Let us first consider how the nuclear reaction rates depend on VPCs. We may write for the general case of two body reactions between incoming particles $i$ and $j$ resulting in outgoing particle $k$ and $l$ [53]

$$i + j \leftrightarrow k + l. \tag{20}$$

The forward reaction rate is defined as

$$f_{ij \to kl} = \frac{n_i n_j}{1 + \delta_{ij}}\langle\sigma v\rangle_{ij \to kl}. \tag{21}$$

Here $n_i$ and $n_j$ are the number densities of the reacting particles, and $\langle\sigma v\rangle_{ij \to kl}$ is the product of the scattering cross section $\sigma$ and the velocity $v$ averaged over the appropriately normalized velocity distribution. The factor with the Kronecker delta $\delta_{ij}$ covers the possibility of $i$ and $j$ being the same type of particles. For three body and higher multibody reactions the equations are more complex. However, our concern is not to determine the reaction rates, but just how they depend on the VPCs. This dependency is determined directly by knowing the dependency of $\langle\sigma v\rangle$ on the VPCs.

A textbook expression for $\langle\sigma v\rangle$ is [54] is

$$\langle\sigma v\rangle = \left(\frac{8}{\pi\mu}\right)^{\frac{1}{2}}\left(\frac{S_0}{(k_B T)^{\frac{3}{2}}}\right)\int_0^\infty e^{-\frac{E}{k_B T}}e^{-\sqrt{\frac{E_G}{E}}}dE. \tag{22}$$

Here $\mu$ is the reduced mass of the particles involved, $S_0$ is related to the S-factor which is determined mostly through accelerator experiments, $T$ is the temperature of the reaction, $E$ is the particle energy and $E_G$ is the Gamow energy. The expression for the Gamow energy is

$$E_G = (\pi\alpha Z_i Z_j)^2 2\mu c^2, \text{ with } \alpha \equiv \frac{1}{4\pi\epsilon_0}\left(\frac{e^2}{\hbar c}\right). \tag{23}$$

Here $\alpha$ is the fine structure constant with $\epsilon_0$ as the permittivity of free space and e the electron charge, and $Z_i$ and $Z_j$ are the atomic numbers of the reacting nuclei.

The integrand in Eq. (22) is approximated by a Gaussian shape with a peak at $E_0 = (k_B T/2)^{2/3}E_G^{1/3}$ that has almost zero value at $E = 0$. Therefore the lower limit on the integral in Eq. (22) can be changed to $-\infty$ without influencing the result and can then be analytically integrated. After integrating we may write the equation

$$\langle\sigma v\rangle = \frac{2^{\frac{5}{3}}\sqrt{2}}{\sqrt{3\mu}}S_0\left(\frac{E_G^{\frac{1}{6}}}{(k_B T)^{\frac{2}{3}}}\right)\exp\left[-3\left(\frac{E_G}{4k_B T}\right)^{\frac{1}{3}}\right]. \tag{24}$$

Let us now see how the parameters in Eq. (24) vary. The Gamow energy varies as $E_G \sim c^2 \sim e^{-2}$ since $\alpha$ is constant due to $\epsilon_0 \sim c^{-2} \sim e^2$. The thermal energy $k_B T$ represents the random motion of the particles and thus relates to the mean of their velocity squared, which can be expressed in terms of the speed of light.

Therefore, $k_B T \sim c^2 \sim e^{-2}$. Thus the factor in parentheses would scales as $e^{-1/3} \times e^{4/3}$ i.e. as $e$. Now the expression in the bracket does not vary since variations of $E_G$ and $k_B T$ cancel each other. What we do not know is how $S_0$ varies. Since $S_0$ has the dimensions of energy times area, and length scales as $e^{-1}$ [46], on pure dimensional ground $S_0$ would appear to scale as $e^{-4}$. However, nuclear cross sections reflect reaction details and is rather complicated for it to be represented by simple Coulomb potential and quantum tunneling represented by Eq. [24]. Moreover, the number density product $n_i n_j$ in Eq. (21) for the reaction rates would scale as $e^6$. The cumulative effect would apparently be for the reaction rate $f_{ij \to kl}$ to scale as $e^3$. Therefore, as an ansatz we adopt a simple parametric form for the reaction rate and determine constraint on $p$ in Eq. (25) below from the observations on helium ($^4$He) abundance:

$$f_{ij \to kl} \sim e^p e. \tag{25}$$

Since the abundance is a linear function of the reaction rates [47], we can effectively include its reaction rate dependence on the variation of physical constants by multiplying the abundances calculated using AlterBBN by $e^{p+1}$.

The neutron lifetime plays a critical role in the nucleosynthesis. So we need to consider how it will be affected by the reduced value of the constants in the BBN epoch. The theory of neutron life time $\tau_n$ [55] is related to the theory of the beta decay [56]. In addition, it appears that there is some dependency of the lifetime on the plasma medium of the early universe through Pauli-blocking of decay electrons and neutrinos by plasma which results in the neutrons living longer [57]. All this makes it not only difficult but also uncertain to reliably determine the dependency on VPCs theory of all the parameters involved in the neutron lifetime theory, and consequently to properly capture the dependency of $\tau_n$ on the VPCs. However, many constants remain invariant against the variation of physical constants, such as the fine structure constant, the Rydberg constant and the Stephan-Boltzmann constant [46], Thomson scattering cross section, as well as the orbital periods of planetary bodies and binary pulsars [58]. It appears to be prudent to assume the same to be true for $\tau_n$ unless it is proven otherwise.

## III. RESULTS

We will now test the VPC model against observed BBN abundances. For this we have modified the AlterBBN code [47] developed for the standard ΛCDM model to make it compliant with the VPC model. The significant input parameters used for the VPC model are $\eta = 6.1 \times 10^{-10}$, number of neutrino species 3.046, neutron lifetime 880.2, the same as for the standard ΛCDM model. All the modifications we have made are presented in Appendix A.

We have calculated abundances of the important light nuclei to see how they vary with the baryon to photon ratio $\eta$. The results are graphically presented in Figure 1. We calculated the abundances for equally spaced $\eta$ values from $1 \times 10^{-10}$ to $1 \times 10^{-9}$, and connected the points with smoothing lines. Each of the abundances has two lines corresponding to the potential uncertainty in the variation of reaction rates constrained by $^4$He abundance uncertainty. The double lines in the stacked plot show abundances of $^4$He, D, $^3$He, and $^7$Li as functions of $\eta$ using the VPC model and single lines show the same for the standard model.

The vertical band represents the cosmic microwave background (CMB) determined baryon to photon ratio with its approximate errors and uncertainties determined by Planck collaboration [59]. The horizontal bands represent the observed abundances and their variation and errors among various currently accepted observations. The plot in Figure 1 is slightly different from the BBN Schramm plot [e.g. 5], but we believe it displays the analysis of our findings more explicitly.

Cyburt et al. [5] have discussed the helium abundance data available from several authors. With errors taken into consideration, the helium abundance $Y_p$ ranges from less than 0.24 to above 0.25. We have therefore conservatively drawn the top horizontal band for $Y_p$ from 0.24 to about 0.25. The same reference also cites that the uncertainty of measurement of the deuterium abundance $D/H$ from various observations over the past twenty years is spread over $(1-4) \times 10^{-5}$. However, as discussed below, the $D/H$ spread has now been narrowed down to much smaller value. This is represented by the second horizontal band from the top. $^3$He/H [60,61] ranges over $(1-4) \times 10^{-5}$ depending on the method of measurement and its source. There is contraint on (D+$^3$He)/H (as discussed below) which suggest $^3$He/H spread is much narrower than above, and thus in conflict with our calculated value. This is represented by the third horizontal band from the top.

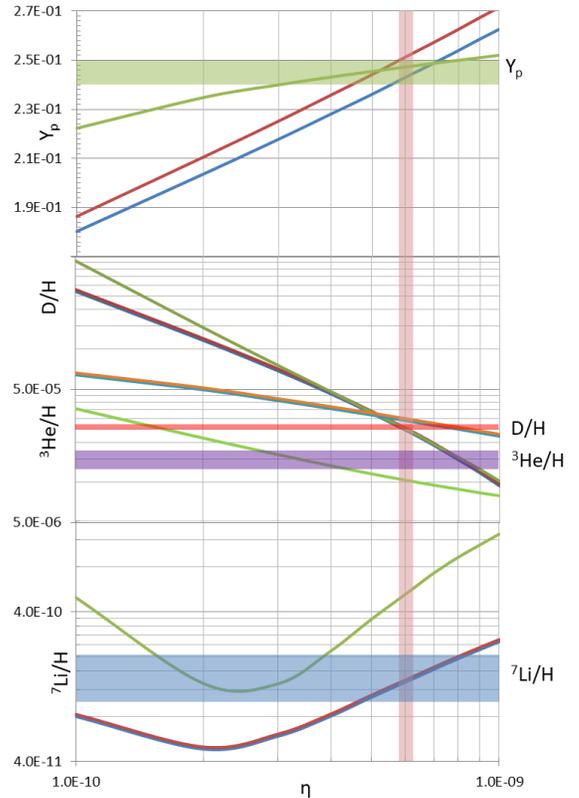

**Figure 1.** This stacked plot shows BBN abundances of $^4$He ($Y_p$), D, $^3$He, and $^7$Li as functions of baryon to photon ratio $\eta$. The double lines are from the VPC model and single lines are from the standard model. Two VPC lines for each element are due to potential uncertainty in the VPC variation of the reaction rates factor constrained by $^4$He abundance. The vertical band represents the cosmic microwave background based $\eta$ with its errors and uncertainties. The horizontal bands represent the observed abundances and their variation and errors among various observations.

Coming now to the bottom horizontal band that represents variations and uncertainties for $^7$Li abundance $^7$Li/H, we have shown it ranging conservatively over $(1-2) \times 10^{-10}$ derived from the review of Tanabashi et al [6]: different analyses, and in some cases different stars and stellar systems, such as in globular clusters,

yield $^7$Li/H as $(1.7 \pm 0.3) \times 10^{-10}$ [62], $(2.19\pm0.28)\times10^{-10}$ [63], and $(1.86\pm0.23)\times10^{-10}$ [64]. The more recent analysis of Sbordone et al. [65] provides $^7$Li/H $= 1.58_{1.30}^{1.93} \times 10^{-10}$. We immediately notice that the theoretical curves for the abundances of the elements considered here are in good agreement with the observations. Obviously, there is no lithium problem.

The actual values of the abundances we computed using the VPC model at $\eta = 6.1 \times 10^{-10}$ and $\tau_n = 880.2\ s$ are as follows: $Y_p= 0.2478_{0.2437}^{0.2520}$, D/H $= 2.453_{2.412}^{2.494} \times 10^{-5}$, $^3$He/H $= 2.940_{2.891}^{2.989} \times 10^{-5}$, and $^7$Li/H $= 1.400_{1.377}^{1.424} \times 10^{-10}$; the higher values are based on $p = -1/6$ and the lower values are based on $p = -1/5$ in Eq. (25). In addition, we also computed $^6$Li/H $= 4.572_{4.496}^{4.648} \times 10^{-15}$ for the $^6$Li abundance and $^7$Be/H $= 1.316_{1.294}^{1.338} \times 10^{-10}$ for the $^7$Be abundance; the former is 44% of the value calculated from the ΛCDM model and the latter only 24%. To our knowledge, there is no reliable observational data to compare them with.

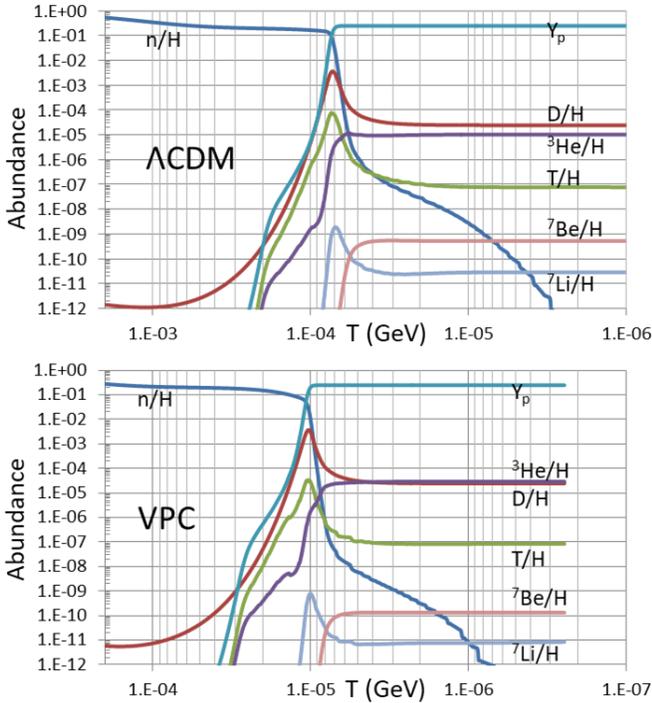

**Figure 2.** The evolution curves of abundances with cosmic temperature for the most abundant nuclides at BBN calculated using AlterBBN code for the ΛCDM and the VPC models.

The evolution curves of abundances against cosmic temperature for the most abundant nuclides at BBN calculated using the AlterBBN code, for the ΛCDM and the VPC models, are presented in Figure 2. The value of $p$ was taken as the average of the two values used in Figure 1, and the baryon to photo ratio $\eta = 6.1 \times 10^{-10}$ and $\tau_n = 880.2\ s$ were used. Apart from the unstable elements $^3$H and $^7$Be decaying into $^3$He and $^7$Li, respectively, all other element abundances may be considered stable until they are influenced by nucleosynthesis in stars. It is easy to notice significant differences between the two plots. One sees that all the VPC curves are shifted with respect to the ΛCDM curves to lower temperature (in GeV) by $e^2$ (=7.389) due to the fact that VPC binding energies are lower by this factor. Other significant difference is in the abundances of $^7$Li and $^7$Be. Both the abundances are significantly lower for the VPC model than for the ΛCDM model. Since the net abundance of $^7$Li is the sum of the two (due to $^7$Be decaying into $^7$Li), the result is that the $^7$Li/H (VPC) is much lower than the $^7$Li/H (ΛCDM). It will be interesting to know which reactions are changed significantly leading to the above and other relatively minor differences. We intend to study it in collaboration with the AlterBBN team and report our findings in a future paper.

## IV. DISCUSSION

BBN is considered a peephole on the earliest universe through the phenomenon of the nucleosynthesis of primordial elements. Most studies assume that such phenomenon is no different than what we study currently in our laboratories. That requires extrapolation over a huge time scale and energy densities that are involved in the BBN. So it is not surprising that BBN findings based on such assumptions are not entirely satisfactory. They rely strongly on the laboratory measured nuclear reaction rates and neutron lifetime. These in turn depend on the assumption that the physical constants are fixed in time. By relaxing this assumption within the constraints of general relativity we have shown that the cosmology up to the scale factor of CMB ($a{\sim}0.001$) can be comfortably explained with the variable physical constants approach [46]. This encouraged us to explore if BBN at $a{\sim}10^{-9}$ is also amenable to this approach.

Results reported in the last section indeed confirm that BBN light element abundances are correctly determined with the varying physical constants approach. It naturally resolves the age old 'lithium problem' without compromising any of the achievements of the standard ΛCDM model. It even retains all the significant parameters used in the BBN code for the standard model including the neutron lifetime, the constraint on the neutrino species, and the baryon to photon ratio.

The primordial deuterium abundance has been extensively studied. The ground based observation of Cooke et al. in 2018 [66] that relates to the O/H absorptions system toward the quasar Q1243+307 yields D/H$= 2.527\ (\pm0.030) \times 10^{-5}$. It should be mentioned that in an earlier paper the same group [67] had determined D/H $= 1.65\ (\pm 0.35) \times 10^{-5}$ from Hubble Space Telescope (HST) observation towards QSO 2206−199. In the same paper they mention other QSO related observations with D/H determined by other authors with values up to $4 \times 10^{-5}$. The observations were further analysed by Pettini et al. in 2008 [68]. Nevertheless, our computed value D/H $= 2.453_{2.412}^{2.494} \times 10^{-5}$ is in reasonable agreement with the most recent and accepted observation of Cooke et al. [66], i.e. $2.527\ (\pm0.030) \times 10^{-5}$.

The $^3$He abundance uncertainty of $(1 - 4) \times 10^{-5}$ [60,61] will reduce to less than $1.5(\pm0.2) \times 10^{-5}$ [77] based on the consideration that the $^3$He abundance should increase with time. The recent discovery of a gradient of $^3$He/H ratio along the Galactocentric radius [78] strongly supports the increasing $^3$He abundance in the Galaxy. Another constraint exists on (D+$^3$He)/H [79,80] from stellar nucleosynthesis and Galactic chemeical evolution theory albeit not directly applicable to BBN. According to our calculation this ratio is $5.393_{5.303}^{5.483} \times 10^{-5}$ whereas it has been measured as $3.6(\pm0.5) \times 10^{-5}$ at the time of solar system formation 4.5 Gyr ago [77]. The current stellar and Galactic chemical evolution theory predicts that (D+$^3$He)/H ratio monotonically increases as a function of time after BBN. Although interstellar D abundance can decrease (via D+P→$^3$He+$\gamma$) conversion to $^3$He, the total sum of D+$^3$He does not changes via this conversion. According to this theory [81] D abundance slowly decreases and $^3$He abundance rather rapidly increases. As a result, (D+$^3$He)/H ratio increases as a function of time. It is currently very difficult to reduce this ratio through stellar nucleosynthesis effects.

Varying physical constants in the context of BBN have been tried by several researchers. We have mentioned some of them in the introduction [4,37-41]. However, in all these studies variation of the constants were determined by fitting the measured light element abundances. The VPC approach is different; in the VPC model the most important parameter $\alpha = 1.8$ is determined analytically [50] and confirmed by fitting the supernovae 1a redshift vs. distance-modulus data [56]. As mentioned above, this model fits SNe Ia data marginally better than the $\Lambda$CDM model, and is consistent with several cosmological observations. Thus, the VPC model is not tailored to satisfy only BBN; it is very general.

It may be meaningful at this stage to discuss the concern of many physicists who insist that dimensionless quantities are the only ones which should be allowed to vary [34,35]. However, there are well known physicists who do not concur with this philosophy. Dirac for example did not agree with Milne who was dimensionless advocate [22]; many more references have been cited in Sec. I. We could think of physical constants belonging to two different categories: constants that are independent of the Hubble expansion of the universe - e.g. the fine structure constant $\alpha$ and proton to electron mass ratio µ; and the constants that may be tied to the Hubble expansion - e.g. the Newton's gravitational constant $G$, and possibly also the speed of light $c$, the Planck's constant $h$ and the Boltzmann constant $k_B$. The variability of any constant can be associated with a dynamical field that evolves due to the action of a background potential which drives the field towards its minimum. Thus, it is implicitly assumed in the fixed constant scenario that the field reached a stable minimum very early in the Big-Bang era resulting in the constancy of the physical constant associated with the field ever since. An evolutionary background potential could therefore cause the field minimum to also evolve. This may explain why all such physical constants might evolve and why studying any constant in isolation might not be prudent.

We have tried to study some constants in the second category and found that in many expressions and formulae they vary in such a way that they cancel the variations of others and thus make their variations unobservable, e.g. the Rydberg constant, the Stephan-Boltzmann constant, and the fine structure constant. The constants in the first category might be varying much more slowly, if varying at all, than those in the second category.

The VPC approach does not impact everyday physics in anyway. Unless the measurement time scales are very large, or measurement precision and temporal sensitivity is very high, well above the noise introduced by perturbations caused by local and astronomical disturbances, it will be hard to directly measure the variations of the constants in a laboratory. The task is made even more difficult as the measuring tools are also evolving; we cannot consider one constant varying and all others fixed in our measurement toolbox. For example, in the determination of $\dot{G}/G$ by lunar laser ranging (LLR) method, one measures the variation of the period of Moon's orbit $P$, and the distance of the Moon $r$ by measuring the time of flight $\tau$ of the laser photons. If we ignore the variation of the speed of light in the measurement of the distance $r = c\tau$, then we get the upper limit on $\dot{G}/G$ several orders of magnitude lower than other methods. The reason for this is the cancellation of the $\dot{G}/G$ by the $3\dot{c}/c$, which can be easily seen by examining the Kepler's 3rd law $P^2 = 4\pi^2 r^3/(GM)$. Taking logarithmic derivative we get $\dot{G}/G = 3\dot{r}/r - 2\dot{P}/P - \dot{M}/M$ [69]. If we substitute $r = c\tau$, we get $\dot{G}/G - 3\dot{c}/c = 3\dot{\tau}/\tau - 2\dot{P}/P - \dot{M}/M$. Since we are in reality measuring $\tau$ and not $r$, we are determining $\dot{G}/G - 3\dot{c}/c$ and not $\dot{G}/G$. But in the VPC approach $\dot{G}/G = 3\dot{c}/c$. This will explain very low value of the upper limit on $\dot{G}/G$ determined by LLR, e.g. by Hofmann & Muller in 2018 [70]. Similar situation may arise in other methods of determining $\dot{G}/G$ when complete or near complete cancellation of the variability of physical constants is not taken into account, such as in asteroseismic method reported by Bellinger & Christensen-Dalsgaardere in 2019 [71], in the BBN method reported by Alveya et al. in 2020 [72], and in the white dwarf cooling method [73].

## V. CONCLUSIONS

We have shown that the varying physical constants approach can naturally resolve the primordial $^7$Li abundance problem without compromising on the deuterium abundance. The value of $^7$Li/H = $1.400^{1.424}_{1.377} \times 10^{-10}$ we have calculated is only about 25% of that estimated using the standard cosmological model, and is consistent with the most agreed observational value of $1.6 (\pm 0.3) \times 10^{-10}$. Although the $^7$Li is as observed, too large a $^3$He overproduction is trigerred. Therefore, the VPC approach cannot satisfactorily explain all the primordial abundances. A more detailed calculation with appropriately accounted S-factors and reaction rates is needed to see if the VPC approach can resolve the $^3$He/H discrepancy. Since the same approach also fitted well the SNe 1a data, and correctly yielded various cosmological observations, including the first peak in the CMB anisotropy power spectrum, we suggest that the VPC approach be tested on various astrophysical observations. We urge that in order to get true VPC dependent results such studies should consider the variation of all the physical constants involved rather than a selected one or few, and also carefully consider how various equations involved are affected due to the time dependency of the physical constants, especially when the scale factor is not very small.

## Acknowledgements

This work has been supported by a generous grant from Macronix Research Corporation. Thanks are due to Madhav Singhal at the University of Western Ontario who made the AlterBBN code operational, and easy to modify and run on the 'cloud'. The author is grateful to Alexandre Arbey of CERN for his continuing support for adapting the code as required for this work, and to Cyril Pitrou of the Institut d'Astrophysique de Paris for providing access to the PRIMAT code with adequate instruction to successfully run it in Mathematica environment. The author wishes to expresses his sincere gratitude to the reviewer of the paper who provided stepwise comments for improving and correcting the paper.

[1] G. Israelian, Nature **489**, 37 (2012).
[2] J. Howk, N. Lehner, B. Fields, et al., Nature **489**, 121 (2012).
[3] V. Singh, J. Lahiri, D. Bhowmick, and D. N. Basu, J. Exp. Theor. Phys. **128**, 707 (2019).
[4] M. T. Clara and C. J. A. P. Martins, A&A, **633**, L11 (2020).
[5] R. H. Cyburt, B. D. Fields, K. A. Olive, and T-H. Yeh, Rev. Mod. Phys. **88**, 015004 (2016).
[6] M. Tanabashi, et al. (Particle Data Group), Phys. Rev. D **98**, 030001 p. 380 (2018).
[7] K. Jedamzik, Phys. Rev. D **70**, 063524 (2004).
[8] Pospelov, M., Pradler, J., 2010, Ann. Rev. Nucl. Part. Sci. 60, 539.
[9] Mathews, G. J., Kedia, A., Sasankan, N., et al., 2020, JPS Conf. Proc. 31, 011033; arXiv: 1909.01245.

[10] Richard, O., Michaud, G., Richer, J., 2005, ApJ., 619, 538.
[11] Fu, X., Bressan, A., Molaro, P., Marigo, P., 2015, MNRAS, 452, 3256.
[12] Lamia, L., Boiano, A., Boiano, C., et al., 2019, ApJ, 879, 23.
[13] Hou, S. Q., He, J. J., Parikh, J. J. A., et al., 2017, ApJ, 834, 165.
[14] Kusakabe, M., Kajino, T., Mathews, G. J., Luo, Y., 2019, PhRvD 99, 043505.
[15] Luo, Y., Kajino, T., Kusakabe, M., Mathews, G. J., 2019, ApJ, 872, 172.
[16] Sasankan, N., Kedia, A., Kusakabe, M., Mathews, G. J., 2018, arXiv:1810.05976.
[17] McDermott, S. D., Turner, M. S., 2018, arXiv:1811.04932.
[18] Thomson W., Tait P. G., 1883, Treatise on natural philosophy, Cambridge University Press.
[19] Weyl H., 1919, Ann. Phys., 59, 129.
[20] Eddington A. S., 1934, New Pathways in Science. Cambridge University Press.
[21] Dirac P. A. M., 1937 Nature, 139, 323.
[22] Dirac P. A. M., 1938, Proc. R. Soc. A, 65, 199.
[23] Barns, C., Dicke R. H., 1961, Phys. Rev., 124, 925.
[24] Einstein A., 1907, Jahrbuch fur Radioaktivitat und Elektronik 4, 11.
[25] Dicke R. H., 1957, Rev. Mod. Phys., 29, 363.
[26] Petit J.-P., 1988, Mod. Phys. Lett. A, 3, 1527; ibid 1733; ibid 2201.
[27] Moffat, J. W., 1993a, Int. J. Mod. Phys. D, 2, 351.
[28] Moffat, J. W., 1993b, Found. Phys., 23, 411.
[29] Albrecht A., Magueijo J., 1999, Phys. Rev. D, 59, 043516.
[30] Barrow, J. D., 1999, Phys. Rev. D, 59, 043515.
[31] Avelino P. P., Martins C. J. A. P., 1999, Phys. Lett. B, 459, 468.
[32] Avelino P. P., Martins C. J. A. P., Rocha G., 2000. Phys. Lett. B, 483, 210.
[33] Moffat, J. W., 2016, Eur. Phys. J. C, 76, 130.
[34] Uzan J.-P., 2003, Rev. Mod. Phys., 75, 403.
[35] Uzan J.-P., 2011, Living Rev. Relativ., 14, 2.
[36] Chiba T., 2011, Prog. Theor. Phys., 126, 993.
[37] Ichikawa, K., Kawasaki, M., 2002, PhRvD 65, 123511.
[38] Landau, S. J., Mercedes E. Mosquera, M. E., Vucetich, H., 2006, ApJ, 637, 38.
[39] Dmitriev, V. F., Flambaum, V. V., Webb, J. K., 2004, Phys. Rev. D 69, 063506.
[40] Coc. Alain, Nunes, N. J., Olive, K. A., Jean-Philippe Uzan, J.-P., Vangioni, E., 2007, Phys. Rev. D 76, 023511.
[41] Berengut, J.C., Flambaum, V. V., Dmitriev, V. F., 2010, Phys. Lett. B 683, 114.
[42] Ellis G. F. R., Uzan J.-P., 2005, Am. J. Phys., 73, 240.
[43] R. P. Gupta, Galaxies **7**, 55 (2019).
[44] Costa R., Cuzinatto R. R., Ferreira E. G. M., Franzmann G., 2019, Int. J. Mod. Phys. D, 28, 1950119.
[45] Franzmann, G., 2017, arXiv:1704.07368.
[46] R. P. Gupta, MNRAS **498**, 4481 (2020).
[47] Arbey, A., Auffinger, J., Hickerson, K. P., Jenssen, E. S., 2018, arXiv:1806.11095.
[48] Pitrou, C., Coc, A., Uzan, J.-P., Vangioni, E., 2018, Phy. Rep., 754, 1.
[49] Maple 2019, Maplesoft, Waterloo, Canada.
[50] R. P. Gupta, Universe **4**, 104 (2018).
[51] Barrow, J. D., Magueijo J., 1999, Phys. Lett. B, 447, 246.
[52] Salzano V., Dabrowski M. P., 2017, ApJ, 851, 97.
[53] Jenssen, E. S., 2016, A Code for Big Bang Nucleosynthesis withLight Dark Matter, Master's Thesis, University of Oslo.
[54] Maoz, D., 2016, Astrophysics in a Nutshell, 2nd Ed. Princeton University Press.
[55] Wietfeldt, E. E., 2018, Atoms, 6, 70.
[56] Nanni, L., 2019, Adv. Stud. Theo. Phys., 13, 281.
[57] Yang, C. T., Birrell, J., Rafelski, J., 2018, arXiv:1805.06543
[58] R. P. Gupta, Orbital timing constraint on $\dot{G}/G$, Phys. Lett. B (under review).
[59] Planck Collaboration, Aghanim N. et al., 2020, Planck 2018 results. I. Overview and cosmological legacy of Planck, A&A 641 A1; arXiv:1807.06025.
[60] Balser, D. S., Bania, T. M., Brockway, C. J., Rood, R. T., Wilson, T. L. 1994, Ap. J., 430, 667.
[61] Olive, K. A., Schramm, D. N., Scully, S. T., Trupan, J. W., 1997, ApJ 479, 752.
[62] Kel'ner, S. R., Kotov, Y. D., 1968, Sov. J. Nucl. Phys. 7, 237.
[63] Kokoulin, R. P., Petrukhin, A. A. in Proceedings of the International Conference on Cosmic Rays, Hobart, Australia, August 16-25, 1971, Vol. 4, p. 2436.
[64] Nikishov, A. I., 1978, Sov. J. Nucl. Phys. 27, 677.
[65] Sbordone, L., Bonifacio, B., Caffau, E., et al., 2010, A&A 522, A26.
[66] Cooke, R. J., Pettini, M, Steidel, C. C., 2018, ApJ, 855, 102.
[67] Pettini, M, Bowen, D. V., 2001, ApJ., 560, 41.
[68] Pettini, M., Zych, B. J., Murphy, M. T., Lewis, A., Steidel, C. C., 2008, MNRAS, 391, 1499.
[69] Merkowitz, S.M., Living Rev. Relativ. 2010, 13, 7.
[70] Hofmann, F.; Müller, J., Class. Quan. Gravity 2018, 35, 035015.
[71] Bellinger, E. P., Christensen-Dalsgaard, J., 2019, , Ap. J. Letts. 887, L1.
[72] Alveya, J., Sabtib, N, Escuderoc, M., Fairbairnd, M., 2020, Eur. Phys. J. C 80, 148.
[73] Benvenuto, O. G., Althaus, L. G., Torres, D. F., 1999, MNRAS, 305, 905.
[74] M. Kusakabe and M. Kawasaki, 2015, MNRAS, 446, 1597.
[75] N. Sasankan, A. Kedia, M. Kusakabe, and G. J. Mathews 2020, Phys. Rev. D. 101, 123532.
[76] K. Mori and M. Kusakabe, 2019, Phys. Rev. D, 99, 083013.
[77] J. Geiss and G. Gloeckler, 1998, Space Sci. Rev. 84, 239.
[78] D. S. Balser and T. M. Bania, 2018, Astron. J. 156, 280.
[79] G. Steigman and M. Tosi, 1995, Ap. J. 453, 173.
[80] F. Iocco, G. Mangano, G. Miele, O. Pisanti and P. D. Serpico , 2009, Phys. Rep. 472, 1.
[81] N. Lagarde, D. Romano, C. Charbonnel, M. Tosi, C. Chiappini and F. Matteucci, 2012, Astron. & Astrophys. 542, A62.

# APPENDIX A
## Modification of the AlterBBN Code

The purpose of this appendix is to show explicitly the changes that we have made to the AlterBBN v.2 code so that the code could represent the VPC model. We have copied the lines directly from the code. Data used are those included in the AlterBBN v.2 publicly available code (Arbey et al. 2018).

*File: src.include.h*

**Original for the standard ΛCDM model:**

```
#define Gn         6.67428e-8  /* in cm^3.g^-1.s^-2 */
#define K_to_eV    8.617330637338339e-05 /* conversion factor T(10**9 K) * K_to_GeV = T(GeV) or T(K) * K_to_eV = T(eV) */
#define m_e        510.9989461e-6 /* electron mass in GeV */
#define g_to_GeV   5.60958884538932e+23 /* conversion factor M(g) * g_to_GeV = M(GeV) */
#define kg_to_GeV  5.60958884538932e+26 /* conversion factor M(kg) * kg_to_GeV = M(GeV) */
#define sigma_SB   0.16449340668482282 /* Stefan-Boltzmann constant = pi^2/60 */
```

```c
#define m_to_GeV     5.067730582705779e+15  /* conversion factor L(m) * m_to_GeV = L(GeV^-1) */
#define cm_to_GeV    5.067730582705779e+13  /* conversion factor L(cm) * cm_to_GeV = L(GeV^-1) */
#define s_to_GeV     1.51926740787113377e+24 /* conversion factor t(s) * s_to_GeV = t(GeV^-1) */
#define G            6.708609142443796e-39  /* Gn*pow(m_to_GeV,3.)*pow(g_to_GeV,-1.)*pow(s_to_GeV,-2.) Newton constant in GeV^-2 */
#define Mplanck      1.2209102930946623e+19 /* in GeV, more precise definition than before */
#define DMpn         0.0012934 /* mass difference between neutron and proton in GeV */
#define zeta         1.6103162253325862 /* 3*k_B/(2*c^2*Mu) in GeV^-1 */
#define k_B          8.617330e-5  /* Boltzmann's constant in GeV/GK */
#define alphaem      0.007297353 /* fine-structure constant */
```

**Modified for the VPC model:**

```c
#define Gn           3.32292834666267E-09  /* VPC(1/e^3) in cm^3.g^-1.s^-2 */
#define K_to_eV      2.46890656372771e-05 /* VPC (1/e^1.25) conversion factor T(10**9 K) * K_to_GeV = T(GeV) or T(K) * K_to_eV = T(eV) */
#define m_e          69.15618710e-6 /* VPC (1/e^2) electron mass in GeV */
#define g_to_GeV     0.759175295231707e+23 /* VPC (1/e^2) conversion factor M(g) * g_to_GeV = M(GeV) */
#define kg_to_GeV    0.75917529523170e+26 /* VPC (1/e^2) conversion factor M(kg) * kg_to_GeV = M(GeV) */
#define sigma_SB     0.16449340668482282 /* VPC (unchanged) Stefan-Boltzmann constant = pi^2/60 */
#define m_to_GeV     37.4457455698794e+15 /* VPC (*e^2) conversion factor L(m) * m_to_GeV = L(GeV^-1) */
#define cm_to_GeV    37.4457455698794e+13 /* VPC (*e^2) conversion factor L(cm) * cm_to_GeV = L(GeV^-1) */
#define s_to_GeV     4.12979698738617e+24 /* VPC (*e) conversion factor t(s) * s_to_GeV = t(GeV^-1) */
#define G            1.34750054524155e-37 /* VPC (*e^3) Gn*pow(cm_to_GeV,3.)*pow(g_to_GeV,-1.)*pow(s_to_GeV,-2.) Newton constant in GeV^-2 */
#define Mplanck      0.27241750391653e+19 /* VPC (1/e^1.5) in GeV, more precise definition than before */
#define DMpn         0.00017504 /* VPC (1/e^2) mass difference between neutron and proton in GeV */
#define zeta         11.898716926 /* VPC (*e^2) 3*k_B/(2*c^2*Mu) in GeV^-1 */
#define k_B          2.468906e-5   /* VPC (1/e^1.25) Boltzmann's constant in GeV/GK */
#define alphaem      0.007297353 /* VPC (unchanged) fine-structure constant */
```

*File: src.bbn.c*

**Original for standard ΛCDM model:**

```c
q9[i]=reacparam[i][9]
sum_DeltaMdY_dt+=Dm[i]/1000.*dY_dt[i]/(M_u*g_to_GeV);
reacparam[12][9]
```

**Modified for VPC model:**

```c
q9[i]=reacparam[i][9]/2.117; //VPC: mass to eV to T > 1/e^0.75=1/2.117
sum_DeltaMdY_dt+=Dm[i]/7389.*dY_dt[i]/(M_u*g_to_GeV); //Dm is in MeV; VPC: 1000 replaced by 7389 - Dm*c^2=Dmc_0^2/e^2 with e^2=7.389
reacparam[12][9]/2.117 //VPC: mass to eV to T > 1/e^0.75=1/2.117 – at 3 places
```

*File: src.cosmodel.c*

**Original for standard ΛCDM model:**
```c
paramrelic->life_neutron=880.2; // Neutron lifetime (PDG2018) 880.2 s
    paramrelic->life_neutron_error=1.; // Neutron lifetime uncertainty (PDG2017) 1 s
double rhorad=pi*pi/30.*geffT*pow(T,4.);
 double rho_photon_1MeV=pi*pi/15.*1.e-12;
```

**Modified for VPC model:**
```c
paramrelic->life_neutron=880.2; // Neutron lifetime (PDG2018) 880.2 s VPC (*e^0.0)
    paramrelic->life_neutron_error=1.; // Neutron lifetime uncertainty (PDG2017) 1 s VPC 1 s
double rhorad=pi*pi/30.*2.71828*geffT*pow(T,4.);// VPC (*e)
double rho_photon_1MeV=pi*pi/15.*2.71828*1.e-12;// VPC (*e)
```

*File: src.bbnrate.c*

**Original for standard ΛCDM model:**
```c
double q = 1.29333217e-3/m_e; /// q=(mn-mp)/me;
```

**Modified for VPC model:**
```c
double q = 1.29333217e-3/7.3890561/m_e; /// q=(mn-mp)/me; VPC 1/e^2
```